# On the inverse problem of the earthquake source and the three-phase relaxation mode of the source after the formation of a main rupture in it

A.V. Guglielmi

*Schmidt Institute of Physics of the Earth, Russian Academy of Sciences; Bol'shaya Gruzinskaya str., 10, bld. 1, Moscow, 123242 Russia; guglielmi@mail.ru (A.G.)*

**Abstract**

The fundamentals of the phenomenological theory of aftershocks are presented. The theory contains an original concept of the proper time of the earthquake source, the course of which, generally speaking, differs from the course of world time. Within the framework of the theory, a new method for processing and analyzing earthquakes has been developed. Analysis of the Tohoku earthquake demonstrated the effectiveness of the method. Three phases of relaxation of the source were discovered. In the initial phase, the deactivation coefficient, which characterizes the source as a dynamic system, is equal to zero. At the end of the initial phase, the deactivation coefficient suddenly acquires a positive value, which remains unchanged throughout the main relaxation phase. The transition from the main phase to the recovery phase is accompanied by a jump in the time derivative of the deactivation coefficient. The recovery phase differs from the initial and main phases by chaotic variations in the deactivation coefficient.

## 1. Introduction

We imagine the earthquake's source as a stress-strain rock mass in which a main rupture spontaneously arises. This main rupture manifests itself as the main shock of the earthquake. The main shock is usually followed by a long-lasting series of aftershocks. Aftershocks form a discrete sequence of underground tremors that reflect the relaxation process of the source after the formation of a main rupture in the continuity of rocks. The evolution of aftershocks is usually described by a continuous function $n(t)$, which is called the instantaneous aftershock frequency. The value of $n(t)$ is calculated by averaging over small time intervals containing a sufficiently large number of events.

At the end of the last century, Omori discovered that the frequency of aftershocks is satisfactorily approximated by the function $n(t) = k/(c+t)$, where $k > 0$, $c > 0$, $t \geq 0$ [1]. The specified dependence is widely known in the literature as Omori's law. It is quite clear that Omori's law is a solution to a certain dynamic equation describing the evolution of the source. It is easy to guess that the equation has the form [2]

$$\frac{dn}{dt} + \sigma n^2 = 0. \qquad (1)$$

Indeed, let us find the solution to the equation of evolution

$$n(t) = \frac{n_0}{1 + n_0 \tau(t)}, \qquad (2)$$

where

$$\tau(t) = \int_0^t \sigma(t) dt. \qquad (3)$$

Here $n_0 = n(0)$ is the initial condition. We will call the value of $\sigma$ the deactivation coefficient, and the value of $\tau$ the proper time of the source. From (2) for $\sigma = \text{const}$ the Omori law follows with accuracy up to the notation.

So far we have not presented anything particularly noteworthy except that aftershock activity on average decreases hyperbolically with the proper time of the source. The difference between proper time and world time is connected with the inconstancy of the phenomenological parameter $\sigma(t)$ of our theory. The dependence of the deactivation coefficient on time cannot be excluded from a priori considerations, since relaxation of the source almost certainly occurs under non-stationary conditions.

The radical advantage of the dynamic equation (1) compared to the Omori law becomes obvious when formulating the inverse problem of the earthquake source [3]. The inverse problem is formulated as follows: Find the deactivation coefficient based on the observed aftershock frequency data. To solve this, we move to a new dependent variable in (1): $n \to g = 1/n$. The formal solution $\sigma = dg'dt$ is usually unstable in practice. Regularization of the inverse problem is reduced to optimal smoothing of the auxiliary function $g(t)$. Let us introduce the notation $T = \langle g \rangle$. Here the angle brackets denote the smoothing of the function $g(t)$. After this, the correct solution to the inverse problem takes the form

$$\sigma(t) = \frac{d}{dt} T(t). \tag{4}$$

Together with B.I. Klain, A.D. Zavyalov and O.D. Zotov, the author has been experimentally studying the relaxation of earthquake sources for many years by analyzing changes in the deactivation coefficient over time. As a result, a number of previously unknown properties and patterns of earthquake source dynamics were discovered (see, for example, review papers [4–6]). In this paper, we focus on the phenomenon of three-phase relaxation of the earthquake source.

## 2. Discreteness and continuity

Let us pay attention to the continuity of the proper time (3), synchronized with the world time by calculating the deactivation coefficient according to formula (4). The averaging and smoothing procedures leveled out the discreteness of the aftershocks, which served as the initial source of information for the formation of the phenomenological theory of relaxation of the source, within the framework of which formula (4) was obtained. The idea arose to try to use the concept of discrete proper time [7, 8]. We are fully aware that, at the current level of knowledge, world time is a continuum, but the proper time of the source is, in one way or another, an artificial concept and manipulation with it is permissible up to a certain limit.

Let us imagine a sequence of underground tremors as the ticking of an imaginary underground clock, the course of which is in some way coordinated with the course of non-stationary dynamic processes in the earthquake source. Our proposal is motivated by the hope that with discrete chronometry of earthquakes we will be able to discover patterns that are not noticeable with standard chronometry based on clocks that keep world time, or when ordering events according to continuous proper time $\tau$.

For the sake of clarity, we will continue to talk about aftershocks, although the proposed concept can also be used in relation to foreshocks. Let us number the sequence of aftershocks with natural numbers $k$ = 1, 2, 3, .. and let us consider the dimensionless unit of proper time to be the number 1, regardless of how large the interval of universal time is between adjacent aftershocks. Let us perform synchronization by assigning to each aftershock the time of its occurrence $t_k$, indicated in the USGS/NEIC catalog. Let us introduce the coordinate plane $(x, t)$ and plot points $t_k$ on it. Here the letter $x$ comes from the word χρόνος.

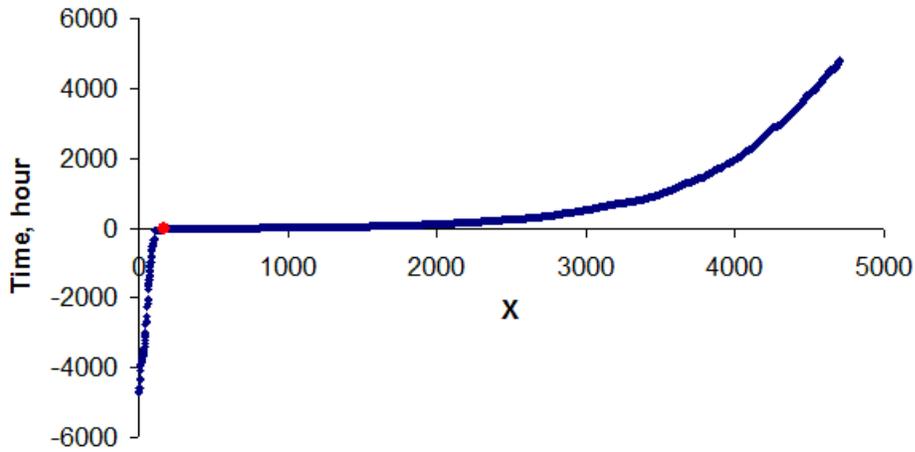

**Fig. 1**. Dependence of the occurrence times of foreshocks, mainshock and aftershocks according to universal time (vertical axis) on the proper time of the Tohoku earthquake source (horizontal axis). The moment of the main shock is marked by a red dot [7].

To illustrate the procedures described above, consider the Tohoku earthquake that occurred on March 11, 2011, with a mainshock magnitude of $M$ = 9.1. In Figure 1, points $t_k$ can be approximated by a piecewise smooth continuous function $t(x)$ and then the deactivation coefficient

$$\sigma = \frac{d}{dx} \ln T. \qquad (5)$$

can be calculated. Here $T = dt/dx$ is the average world time interval between adjacent aftershocks.

Thus, if we order the aftershocks by continuous proper time, then the relaxation equation of the source and its solution have the form

$$\frac{dn}{d\tau} + n^2 = 0, \quad n = \frac{n_0}{1 + n_0 \tau}. \qquad (6)$$

If we order the aftershocks by discrete proper time, then

$$\frac{dn}{dx} + \sigma n = 0, \quad n = n_0 \exp\left(-\int_0^x \sigma(x')dx'\right). \tag{7}$$

In the first case, the evolution equation is nonlinear. The aftershock frequency decreases hyperbolically with time. In the second case, the equation is linear and the aftershock frequency decreases exponentially over time. The analysis of Figure 1, carried out in [7], convincingly confirmed our conclusion about the exponential decrease in aftershock activity when chronometry is based on the discrete proper time of the source.

### 3. Relaxation mode of the source

It is useful to look at source relaxation from two perspectives. Let us first consider the variation of the deactivation coefficient when ordering aftershocks by discrete time, and then when ordering by continuous time.

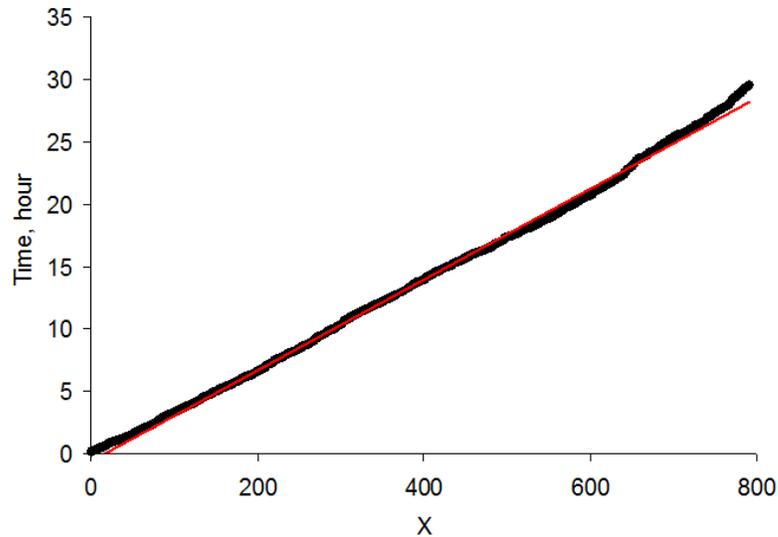

**Fig. 2**. Linear aproximation of a subset of points $k = 1, 2, .., 800$. The main shock occurred at $t = 0$.

In Figure 1, aftershocks are represented by a set of 4537 points. For obvious reasons, we see a solid curve, but in fact all the points are isolated from each other. A larger scale analysis of the dataset suggests that the first 800 points should be approximated by a linear function, as shown in Figure 2. Subsequent points are well approximated by an exponential function. With a linear dependence of $t$ on $x$, the average interval between adjacent aftershocks $T = $ const. Therefore, $\sigma = 0$ in accordance with formula (5). With an exponential dependence of $t$ on $x$, the deactivation coefficient is non-zero. For the Tohoku earthquake, calculation using formula (5) yields a value of $\sigma = 0.0014$. Thus, we found a sharp change in the relaxation regime at $x = 800$. A jump in the phenomenological parameter characterizing the source as a dynamic system indicates a critical transition reminiscent of a bifurcation.

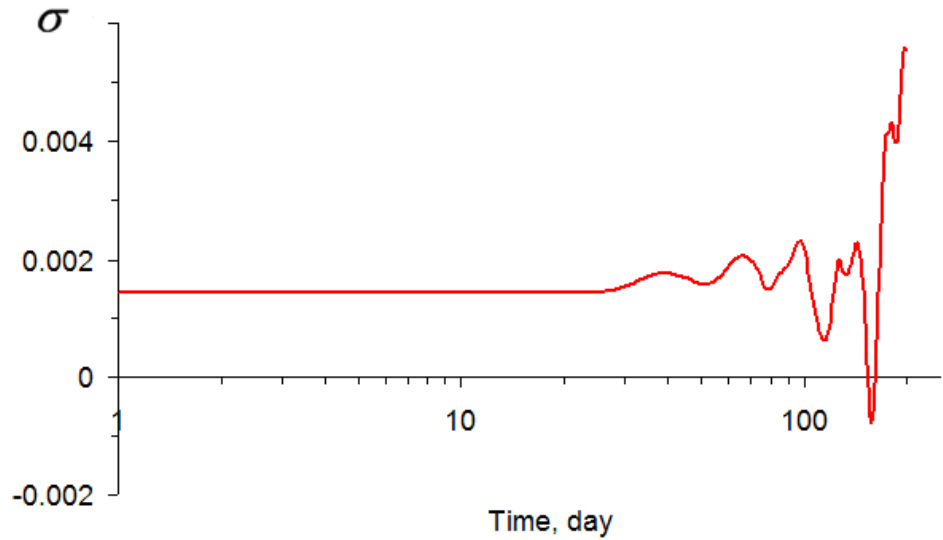

**Fig. 3**. Variation of the deactivation coefficient of the Tohoku earthquake source.

Now let's see what the evolution of the source looks like when the aftershocks are ordered over continuous time. Figure 3 shows the result of calculating the deactivation coefficient using formula (4). The short initial relaxation phase was not reflected in the graph because we performed a smoothing procedure for the auxiliary function $g(t)$. At the end of the initial phase, the deactivation coefficient increased abruptly by $\Delta\sigma = 0.0014$ and the next phase began, which we will call the main phase of relaxation of the source.

The main phase is characterized by the constancy of the deactivation coefficient. It can be called the *Omori epoch*, since at $\sigma = \text{const}$ the classical Omori law is strictly fulfilled. During the Omori epoch, 2,200 aftershocks occurred. (Recall that 800 aftershocks occurred during the initial phase.)

The Omori epoch lasted 20 days, after which the second phase transition occurred and the third, restorative phase of relaxation began, which lasted for six months. During the recovery phase, 1,537 aftershocks occurred. At the second bifurcation, the function $\sigma(t)$ remains continuous, but its derivative experiences a finite jump, characteristic of second-order phase transitions.

Let us emphasize the radical difference between the main phase and the recovery phase. In the main phase $\sigma = \text{const}$, evolution proceeds in a regular manner and the average frequency of aftershocks decreases over time in a completely predictable manner. In contrast, during the recovery phase, the $\sigma(t)$ function is chaotic and aftershock activity is unpredictable.

## 4. On the foreshocks

The main shock of an earthquake is sometimes preceded by foreshocks. For example, 166 tremors were observed in the 200 days before the Tohoku earthquake. It is possible to order these foreshocks by discrete proper time, similar to how it is done for aftershocks.

In Figure 1, the foreshocks are located to the left of the red dot. The set of 166 discrete values of $t_k$, plotted on the coordinate plane, is approximated by the linear function $t(x) = 32x - 4169$ with high accuracy. This means that the mean universal time interval $T = dt/dx$ between two adjacent foreshocks remains unchanged over a very long observation period.

It is interesting to note that the first 800 aftershocks in the initial relaxation phase are also approximated by a linear function, as we indicated above. A sharp increase in the slope of the aftershock line to the horizontal axis on the coordinate plane compared to the foreshock line is associated with a significant increase in the frequency of tremors after the formation of a main rupture in the continuity of rocks at the source.

## 5. Discussion

When analyzing aftershocks, instead of the Omori law, the Hirano-Utsu law $n(t) = k/(c+t)^p$ is usually used, where $p > 0$ (see, for example, [9–17]). In this regard, it must be said that Omori's law, while not holistic, is strictly fulfilled in the main phase of relaxation of the source. In contrast, the Hirano-Utsu law is not applicable to describing the evolution of aftershocks in any of the three relaxation phases of the source. Indeed, it is easy to verify that the Hirano-Utsu law for $p > 1$ predicts a monotonic increase, and for $p < 1$ a monotonic decrease in the deactivation coefficient over time. This property is not observed in the initial phase of relaxation, when the deactivation coefficient is zero. In the main phase, the Omori law is fulfilled, and in the recovery phase, the deactivation coefficient is a non-monotonic function of time.

We see that the phenomenological theory of aftershocks presented in reviews [4–6] contains tools suitable for a fine analysis of the dynamics of an earthquake source. We focused on detecting three phases of source relaxation. Let us point out another surprising result of the experimental study, carried out taking into account the phenomenological theory of earthquakes. When studying the spatiotemporal distribution of underground shocks, the epicenters of foreshocks and aftershocks were ordered according to the discrete proper time of the source. As a result, the phenomenon of foreshock convergence and aftershock divergence was discovered [6]. If the epicenters of foreshocks and aftershocks are considered as tracers visualizing the Umov-Poynting energy flows in the source, then the phenomenon can be interpreted as indirect evidence of the influx of energy to the hypocenter before the main shock and the outflow of energy from the hypocenter after the main shock.

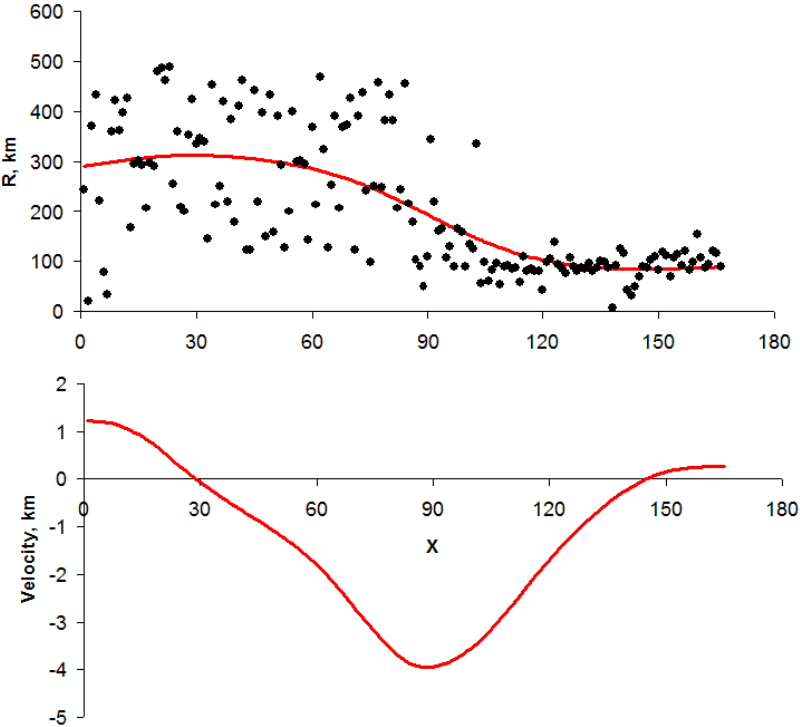

**Fig. 4**. Distances of aftershock epicenters from the Tohoku mainshock epicenter (top) and average foreshock convergence velocity (bottom) as a function of proper time.

We will conclude the discussion by estimating the rate of foreshock convergence

$$V = dR/dt = U/T. \qquad (8)$$

Here $R(x)$ is a smooth function approximating the change in the distance of foreshock epicenters from the epicenter of the main shock, $U = dR/dx$, $T = dt/dx$. To estimate $V$ we will use Figure 4. In the top panel, the red line $R(x)$ approximates the experimental points. On the bottom panel we see a kind of "velocity bay". The maximum absolute velocity is $|U| = 4$ km. According to [8] $T = 47$ h at $x = 90$. Therefore, the maximum convergence velocity is $|V| = 85$ m/h.

The question of whether the velocity bay could be used for medium-term forecasting of the Tohoku earthquake is, of course, unanswered at this time. To study the stability of the indicated anomaly in the convergence rate of foreshocks, it is necessary to analyze a sufficiently large number of catastrophic earthquakes.

## 6. Conclusion

Within the framework of the phenomenological theory of aftershocks, the foundations of which are laid in the works [2–6], the inverse problem of the source was solved and, using the example of the Tohoku earthquake, three phases of relaxation of the source after the formation of a main rupture in the continuity of rocks were discovered. In the initial phase, the deactivation coefficient, which characterizes the state of the source as a dynamic system, is equal to zero. At the end of the initial phase, the deactivation coefficient experiences a jump and acquires a final positive value, which remains unchanged throughout the main relaxation phase. The transition from the main phase to the third phase, called the recovery phase, is accompanied by a jump in the time derivative of the deactivation coefficient. The recovery phase differs from the initial and main phases by chaotic variations in the deactivation coefficient. The obtained results suggest that at the end of the initial phase, a first-order phase transition occurs in the source, and at the end of the main phase, a second-order transition occurs.

*Acknowledgments*. I express my sincere gratitude to B.I. Klain, A.D. Zavyalov and O.D. Zotov for fruitful discussions and support. Together with them, a phenomenological theory of aftershocks was developed. I would like to express my special gratitude to O.D. Zotov, who prepared the graphic series for this paper. I thank colleagues at the US Geological Survey for lending us their earthquake catalogs USGS/NEIC for use. The work was carried out within the framework of the planned tasks of the Ministry of Science and Higher Education of the Russian Federation to the Institute of Physics of the Earth of the Russian Academy of Sciences.

*About the author*. 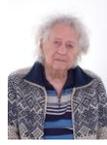 Guglielmi Anatol Vladimirovich was born on October 27, 1935 in the village of Novo-Tavolzhanka, Shebekinsky district, Kursk region. In 1959 he graduated from the Radiophysics Department of Gorkovsky (Nizhny Novgorod) State University, qualified as a research physicist. Teachers: Boris Nikolaevich Gershman and Vitaly Lazarevich Ginzburg. Academic degree: Doctor of Physical and Mathematical Sciences. Academic title: professor of geophysics. Current position: Chief Researcher, Institute of Physics of the Earth, Russian Academy of Sciences. Research interests: cosmical electrodynamics and earthquake physics.

**References**

1. *Omori F.* On the aftershocks of earthquake // J. Coll. Sci. Imp. Univ. Tokyo. 1894. V. 7. P. 111–200.

2. *Guglielmi A.V.* Interpretation of the Omori Law // Izv. Phys. Solid Evarth. 2016. V. 52. P. 785–786 // arXiv:1604.07017 [physics.geo-ph].

3. *Guglielmi A.V.* Omori's law: a note on the history of geophysics // Phys. Usp. 2017. V. 60. P. 319–324. DOI: 10.3367/UFNe.2017.01.038039.

4. *Zavyalov A., Zotov O., Guglielmi A., Klain B.* On he Omori Law in the physics of earthquakes // Appl. Sci. 2022, vol. 12, issue 19, 9965. https://doi.org/10.3390/app12199965

5. *Guglielmi, A.V., Klain, B.I., Zavyalov, A.D., and Zotov, O.D.* The fundamentals of a phenomenological theory of earthquakes // J. Volcanol. Seismol., 2023, vol. 17, no. 5, pp. 428–438.

6. *Guglielmia A.V., Zavyalov A.D., Zotov O.D., Klain B.I.* A method for processing and analysis of aftershocks due to a tectonic earthquake: A new look at an old problem //

J. Volcanology and Seismology. 2025. V. 19. No. 2. P. 196–202. DOI: 10.1134/S0742046325700022.

7. *Zotov O.D., Guglielmi A.V.* Evolution of the Tohoku earthquake aftershocks in the framework of the phenomenological theory of aftershocks // arXiv:2508.04323 [physics.geo-ph].

8. *Guglielmi A.V., Zotov O.D.* Relativity of Time in Earthquake Physics // arXiv:2509.04858 [physics.geo-ph].

9. *Hirano R.* Investigation of aftershocks of the Great Kanto earthquake at Kumagaya // Kishoshushi. Ser. 2. 1924. V. 2. P. 77–83 (in Japanese).

10. *Jeffreys H.* Aftershocks and periodicity in earthquakes // Gerlands Beitr. Geophys. 1938. V. 56. P. 111–139.

11. *Utsu T.* Magnitudes of earthquakes and occurrence of their aftershocks // Zisin Ser. 2, 10, 35-45.

12. *Utsu T., Ogata Y., Matsu'ura R.S.* The centenary of the Omori formula for a decay law of aftershock activity // J. Phys. Earth. 1995. V. 43. № 1. P. 1–33. http://dx.doi.org/10.4294/jpe1952.43.1

13. *Ogata Y., Zhuang J.* Space–time ETAS models and an improved extension // Tectonophysics. 2006. V. 413, Iss. 1–2. P. 13-23.

14. *Faraoni V.* Lagrangian formulation of Omori's law and analogy with the cosmic Big Rip, Eur. Phys. J. C., 2020, 80(5): 445. DOI:10.1140/epjc/s10052-020-8019-2

15. *Rodrigo M.R.* A spatio-temporal analogue of the Omori-Utsu law of aftershock sequences // Cornell University Library: arXiv:2111.02955v1 [physics.geo-ph], submitted on 21 Oct 2021, pp. 1-12. https://doi.org/10.48550/arXiv.2111.02955

16. *Salinas-Martínez A., Perez-Oregon J., Aguilar-Molina A.M., Muñoz-Diosdado A., Angulo-Brown F.* On the possibility of reproducing Utsu's Law for earthquakes with a spring-block SOC model // Entropy. 2023. V. 25. P. 816. https://doi.org/10.3390/e25050816

17. *Bogomolov L.M., Rodkin M.V., Sychev V.N.* Instanton Representation of Foreshock—Aftershock Sequences // Izv. Phys. Solid Evarth. 2025, № 2. P. 43–57. DOI: 10.31857/S0002333725020045,

18. *Guglielmi A.V., Zavyalov A.D., Zotov O.D., Klain B.I.* On Three Laws in Earthquake Physics // Journal of Volcanology and Seismology. 2025. Vol. 19. No. 5. P. 480–489. DOI: 10.1134/S074204632570023X.